\documentclass{llncs}
\usepackage[latin1]{inputenc}
\usepackage{amsmath}
\usepackage{amsfonts}
\usepackage{amssymb}
\usepackage{upgreek}
\renewcommand{\subparagraph}{}
\usepackage[small]{titlesec}
\usepackage{multirow}  
\usepackage{wrapfig}   
\usepackage{graphicx}  
\usepackage{subfig}    
\usepackage[dvips,bookmarks=true,
            plainpages=false,naturalnames=true,
            colorlinks=true,
            linkcolor=blue,citecolor=blue,urlcolor=blue]
        {hyperref}

\title{A Practical Attack on the MIFARE Classic}
\author{Gerhard de Koning Gans, Jaap-Henk Hoepman, and Flavio D. Garcia}
\institute{Institute for Computing and Information Sciences \\
  Radboud University Nijmegen\\
  P.O. Box 9010, 6500 GL \ Nijmegen, The Netherlands\\
  \email{gkoningg@sci.ru.nl},\\\email{jhh@cs.ru.nl},\\\email{flaviog@cs.ru.nl}
}

\begin{document}
\setlength{\abovedisplayskip}{3pt plus 1pt minus 1pt}
\setlength{\belowdisplayskip}{3pt plus 1pt minus 1pt}

\maketitle



\begin{abstract}
The \textsc{mifare} Classic is the most widely used contactless smart card in the market. Its design and implementation details are kept secret by its manufacturer.
This paper studies the architecture of the card and the communication protocol between card and reader. Then it gives a practical, low-cost, attack that recovers secret information from the memory of the card.
Due to a weakness in the pseudo-random generator, we are able to recover the keystream generated by the CRYPTO1 stream cipher. We exploit the malleability of the stream cipher to read \emph{all} memory blocks of the first sector of the card. Moreover, we are able to read \emph{any} sector of the memory of the card, provided that we know \emph{one} memory block within this sector. Finally, and perhaps more damaging, the same holds for \emph{modifying} memory blocks.
\end{abstract}

\section{Introduction}
RFID and contactless smart cards have become pervasive technologies nowadays. Over the last few years, more and more systems adopted this technology as replacement for barcodes, magnetic stripe cards and paper tickets for a variety of applications. Contact-less cards consist of a small piece of memory that can be accessed wirelessly, but unlike RFID tags, they also have some computing capabilities. Most of these cards implement some sort of simple symmetric-key cryptography, which makes them suitable for applications that require access control.

A number of high profile applications make use of contactless smart cards for access control. For example, they are used for payment in several public transport systems like the Octopus card\footnote{\url{http://www.octopuscards.com/}} in Hong Kong, the Oyster card\footnote{\url{http://oyster.tfl.gov.uk}} in London, and the OV-Chipkaart\footnote{\url{http://www.ov-chipkaart.nl/}} in The Netherlands, among others. Many countries have already incorporated a contactless card in their electronic passports~\cite{hoepman_iwsec_2006} and several car manufacturers have it embedded in their car keys as an anti-theft method. Many office buildings and even secured facilities like airports and military bases, use contactless smart cards for access control.

On the one hand, the wireless interface has practical advantages: without mechanical components between readers and cards, the system has lower maintenance costs, is more reliable, and has shorter reading times, providing higher throughput. On the other hand, it represents a potential threat to privacy~\cite{hoepman_iwsec_2006} and it is susceptible to relay, replay and skimming attacks that were not possible before.

There is a huge variety of cards on the market. They differ in size, casing, memory and computing power. They also differ in the security features they provide. A well known and widely used system is \textsc{mifare}. \textsc{mifare} is a product family from NXP semiconductors (formerly Philips). According to NXP there are about 200 million \textsc{mifare} cards in use around the world, covering 85\,\% of the contactless smartcard market. The \textsc{mifare} family contains four different types of cards: Ultralight, Standard, DESFire and SmartMX. The \textsc{mifare} Classic cards come in three different memory sizes: 320B, 1KB and 4KB. The \textsc{mifare} Classic is the most widely used contactless card in the market. Throughout this paper we focus on this card. \textsc{mifare} Classic provides mutual authentication and data secrecy by means of the so called CRYPTO1 stream cipher. This cipher is a proprietary algorithm of NXP and its design is kept secret.

Nohl and Pl\"otz~\cite{CCC} have recently reverse engineered the hardware of the chip and exposed several weaknesses. Among them, due to a weakness on the pseudo-random generator, is the observation that the 32-bit nonces used for authentication have only 16 bits of entropy. They also noticed that the pseudo-random generator is stateless. They claim to have knowledge of the exact encryption algorithm which would facilitate an off-line brute force attack on the 48-bit keys. Such an attack would be feasible, in a reasonable amount of time, especially if dedicated hardware is available.

\paragraph{Our Contribution}

We used a Proxmark III\footnote{\url{http://cq.cx/proxmark3.pl}} to analyze \textsc{mifare} cards and mount an attack.
To do so, we have implemented the ISO 14443-A functionality on the Proxmark, since only ISO 14443-B was implemented at that time. We programmed both processing and generation of reader-to-tag and tag-to-reader communication at physical and higher levels of the protocol. The source code of the firmware is available in the public domain\footnote{\url{http://www.proxmark.org}}. Concurrently, and independently from Nohl and Pl\"otz results, we also noticed a weakness in the pseudo-random generator.

Our contribution is threefold: First and foremost, using the weakness of the pseudo-random generator, and given access to a particular \textsc{mifare} card, we are able to recover the keystream generated by the CRYPTO1 stream cipher, without knowing the encryption key. Secondly, we describe in detail the communication between tag and reader. Finally, we exploit the malleability of the stream cipher to read \emph{all} memory blocks of the first sector (sector zero) of the card (without having access to the secret key). In general, we are able to read \emph{any} sector of the memory of the card, provided that we know \emph{one} memory block within this sector. After eavesdropping a transaction, we are always able to read the first 6 bytes of every block in that sector, and in most cases also the last 6 bytes. This leaves only 4 unrevealed bytes in those blocks.

We would like to stress that we notified NXP of our findings before publishing
our results. Moreover, we gave them the opportunity to discuss with us how to
publish our results without damaging their (and their customers) immediate
interests. They did not take advantage of this offer.

\paragraph{Consequences of our attack}
Any system using \textsc{mifare} Classic cards that relies on the secrecy or
the authenticity of the information stored on sector zero is now insecure.
Our attack recovers, in a few minutes, \emph{all} secret information in
that sector. It also allows us to \emph{modify} any information stored there.
This is also true for most of the data in the remaining sectors, depending on the specific scenario.
Besides, our attack complements Nohl and Pl\"otz results providing the necessary plaintext for a brute force attack on the keys. This is currently work in progress.

%


\paragraph{Outline of this paper}
Section~\ref{mifare} describes the architecture of the \textsc{mifare} cards and the communication protocol.
Section~\ref{hardware} describes the hardware used to mount the attack. Section~\ref{comchar} discusses the protocol by a sample trace. Section~\ref{weakness} exposes weaknesses in the design of the cards. The attack itself is described in Section~\ref{attack}. Finally, Section~\ref{conclusions} gives some concluding remarks and detailed suggestions for improvement.

\section{MIFARE Classic} \label{mifare}
Contactless smartcards are used in many applications nowadays. Contactless cards are based on \textit{radio frequency identification} technology (RFID) \cite{HANDBOOK}. In 1995 NXP, Philips at that time, introduced \textsc{mifare}\footnote{\url{http://www.nxp.com}}. Some target applications of \textsc{mifare} are public transportation, access control and event ticketing. The \textsc{mifare} Classic \cite{NXPMIFARE4K} card is a member of the \textsc{mifare} product family and is compliant with ISO 14443 up to part 3. ISO 14443 part 4 defines the high-level protocol and here the implementation of NXP differs from the standard. Section \ref{communication} discusses the different parts of the ISO standard.

\subsection{Communication Layer} \label{communication}
The communication layer of the \textsc{mifare} Classic card is based on the ISO 14443 standard \cite{ISO14443}. This ISO standard defines the communication for identification cards, contactless integrated circuit(s) cards and proximity cards. The standard consists of four parts.\\
Part 1 describes the physical characteristics and circumstances under which the card should be able to operate.\\
Part 2 defines the communication between the reader and the card and vice versa. The data can be encoded and modulated in two ways, type A and type B. \textsc{mifare} Classic uses type A. For more detailed information about the communication on RFID we refer to the ``RFID Handbook'' by Klaus Finkenzeller~\cite{HANDBOOK}.\\
Part 3 describes the initialization and anticollision protocol. The \textit{anticollision} is needed in order to select a particular card when more cards are present within the reading range of the reader. After a successful initialization and anticollision the card is in an active state and ready to receive a command.\\
Part 4 defines how commands are send. This is the point where \textsc{mifare} Classic differs from the ISO standard, using a proprietary and undisclosed protocol. The \textsc{mifare} Classic starts with an authentication, after that all communication is encrypted. On every eight bits a parity bit is computed to detect transmission errors. In the \textsc{mifare} Classic protocol this parity bit is also encrypted which means that integrity checks are only possible in the application layer.

\subsection{Logical Structure} \label{logicalstructure}
\begin{figure}[htb!]
\begin{center}
\includegraphics[height=60mm]{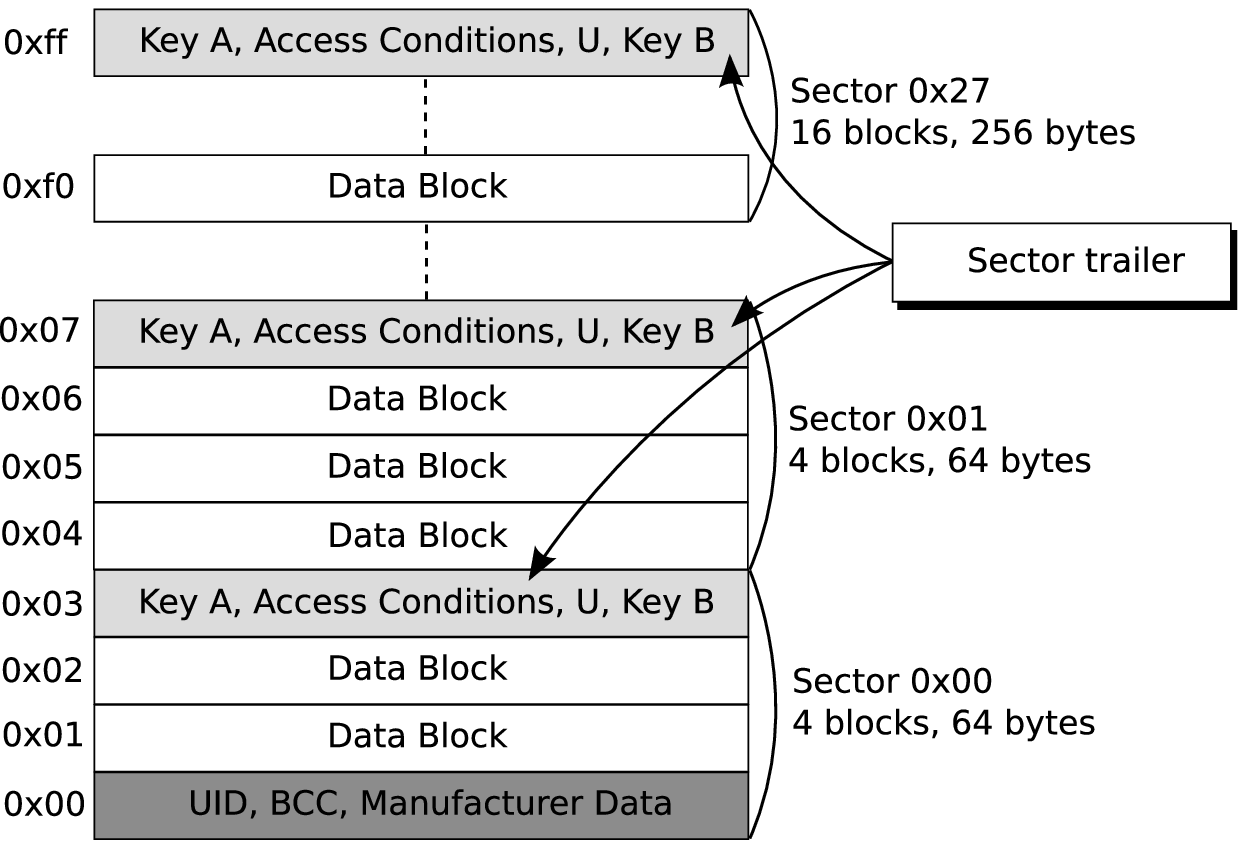}
\caption{\textsc{mifare} Classic 4k Memory}
\vspace{-20pt}
\label{picmemory}
\end{center}
\end{figure}

\begin{figure}[hb]
\centering
\subfloat[Sector Trailer]{\label{picsectortrailer}\includegraphics[width=60mm]{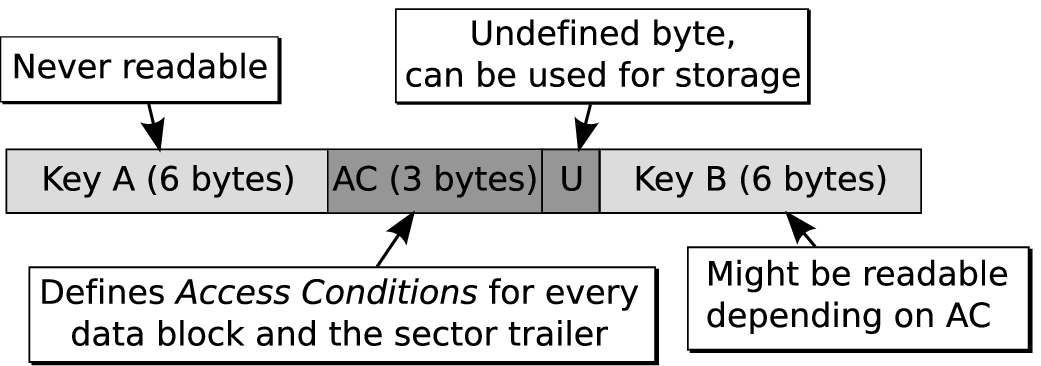}}
\subfloat[Value Block]{\label{picvalueblock}\includegraphics[width=60mm]{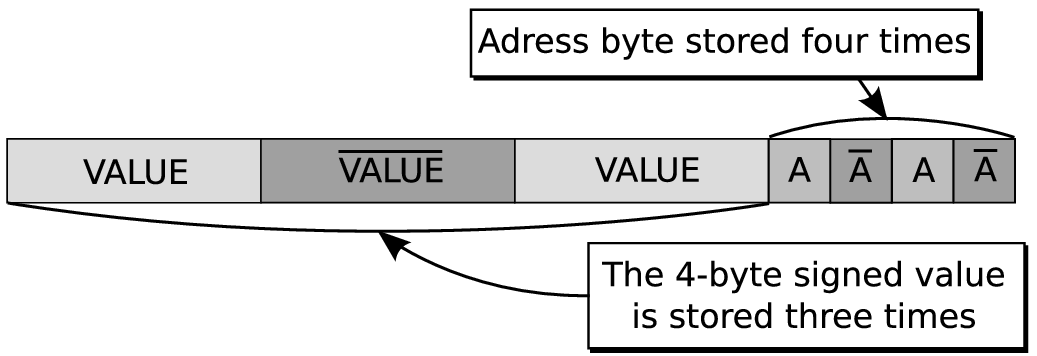}}
\caption{Block contents}
\vspace{-20pt}
\label{picblockcontents}
\end{figure}

A \textsc{mifare} Classic card is in principle a memory card with few extra functionalities. The memory is divided into data blocks of 16 bytes. Those data blocks are grouped into sectors. The \textsc{mifare} Classic 1k card has 16 sectors of 4 data blocks each. The first 32 sectors of a \textsc{mifare} Classic 4k card consists of 4 data blocks and the remaining 8 sectors consist of 16 data blocks. Every last data block of a sector is called \textit{sector trailer}. A schematic of the memory of a \textsc{mifare} Classic 4k card is shown in Figure \ref{picmemory}.
\paragraph{}
Note that block 0 of sector 0 contains special data. The first 4 data bytes contain the unique identifier of the card (UID) followed by its 1-byte \textit{bit count check} (BCC). The bit count check is calculated by successively XOR-ing all UID bytes. The remaining bytes are used to store manufacturer data. This data block is read-only.
The reader needs to authenticate for a sector before any memory operations are allowed. The sector trailer contains the secret keys \textit{A} and \textit{B} which are used for authentication. The \textit{access conditions} define which operations are available for this sector.\\
The sector trailer has special access conditions. Key A is never readable and key B can be configured as readable or not. In that case the memory is just used for data storage and key B cannot be used as an authentication key. Besides the access conditions (AC) and keys, there is one data byte (U) remaining which has no defined purpose. A schematic of the sector trailer is shown in Figure \ref{picsectortrailer}.
A data block is used to store arbitrary data or can be configured as a \textit{value block}. When used as a value block a signed 4-byte value is stored twice non-inverted and once inverted. Inverted here means that every bit of the value is XOR-ed with 1.
These four bytes are stored from the least significant byte on the left to the most significant byte on the right. The four remaining bytes are used to store a 1-byte block address that can be used as a pointer.


\subsection{Commands}
The command set of \textsc{mifare} Classic is small.
Most commands are related to a data block and require the reader to be authenticated for its containing sector. The access conditions are checked every time a command is executed to determine whether it is allowed or not. A block of data might be configured to be read only. Another example of a restriction might be a value block which can only be decremented.
\paragraph{Read and Write}
The read and write commands read or write one data block. This is either a data block or a value block. The write command can be used to format a data block as value block or just store arbitrary data.
\paragraph{Decrement, Increment, Restore and Transfer}
These commands are only allowed on data blocks that are formatted as value blocks. The \textit{increment} and \textit{decrement} commands will increment or decrement a value block with a given value and place the result in a memory register. The \textit{restore} command loads a value into the memory register without any change. Finally the memory register is transferred in the same block or transferred to another block by the \textit{transfer} command.

\subsection{Security Features}
The \textsc{mifare} Classic card has some built-in security features.
The communication is encrypted by the proprietary stream cipher CRYPTO1.

\subsubsection{Keys}
The 48-bit keys used for authentication are stored in the sector trailer of each sector (see section \ref{logicalstructure}). \textsc{mifare} Classic uses symmetric keys.

\subsubsection{Authentication Protocol} \label{authentication}
\textsc{mifare} Classic makes use of a mutual three pass authentication protocol that is based on ISO 9798-2 according to the \textsc{mifare} documentation~\cite{NXPMIFARE4K}. However, it turned out that this is not completely true~\cite{DS2008}. In this paper we only use the first initial nonce that is send by the card. The reader sends a request for sector authentication and the card will respond with a 32-bit nonce $N_{C}$. Then, the reader sends back an 8-byte answer to that nonce which also contains a reader random $N_{R}$. This answer is the first encrypted message after the start of the authentication procedure. Finally, the card sends a 4-byte response. As far as our attack is concerned this description captures all the necessary information.

\section{Hardware and Software} \label{hardware}
An RFID system consists of a transponder (card) and a reader \cite{HANDBOOK}. The reader contains a radio frequency module, a control unit and a coupling element to the card. The card contains a coupling element and a microchip. The control unit of a \textsc{mifare} Classic enabled reader is typically a \textsc{mifare} microchip with a closed design. This microchip communicates with the application software and executes commands from it. Note that the actual modulation of commands is done by this microchip and not by the application software. The design of the microchip of the card is closed and so is the communication protocol between card and reader.

\begin{wrapfigure}{r}{130px}
  \begin{center}
    \includegraphics[width=130px]{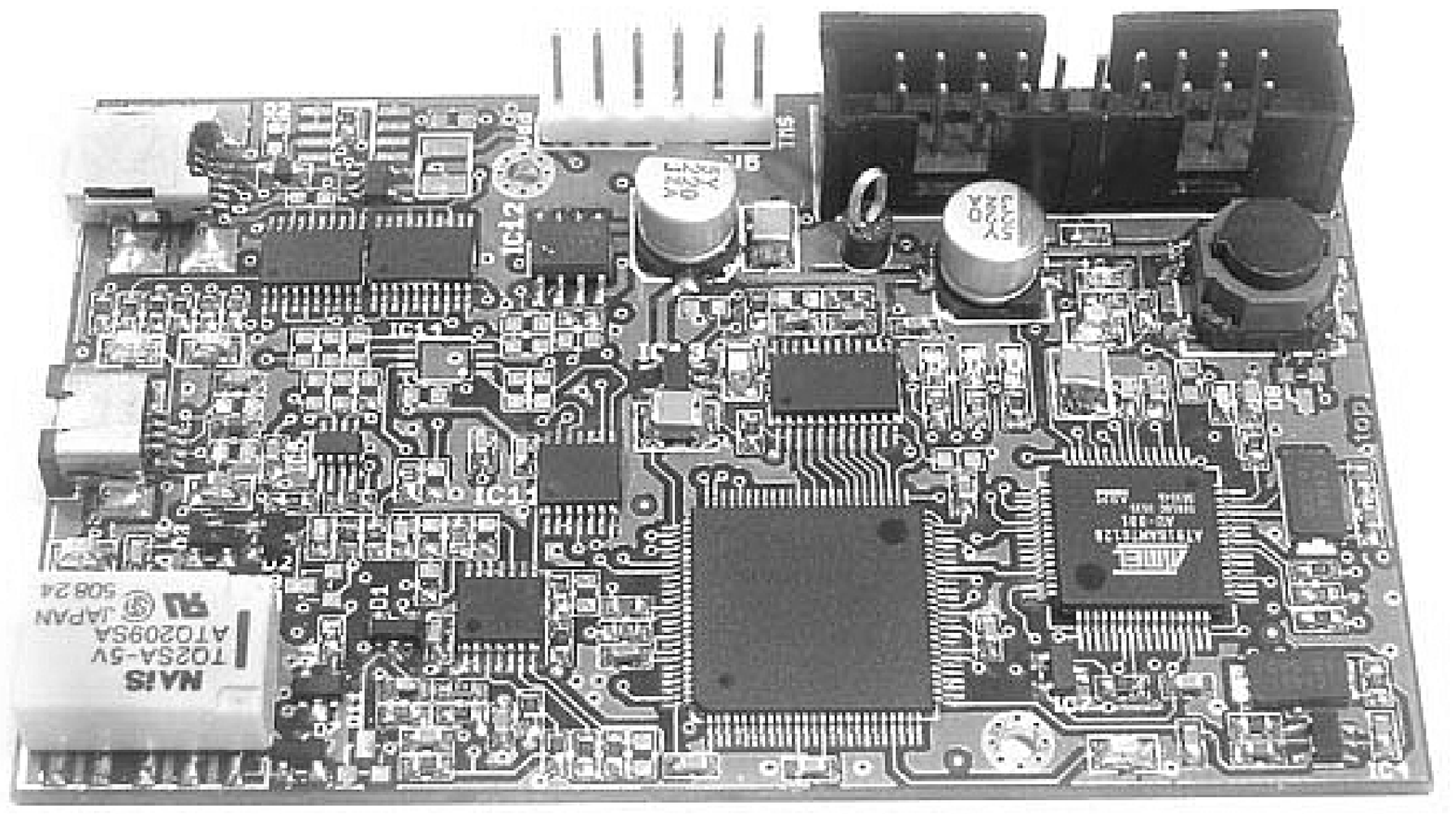}
  \end{center}
  \caption{The Proxmark III}
\end{wrapfigure}

We want to evaluate the security properties of the \textsc{mifare} system.
Therefore we need hardware to eavesdrop a transaction. It should also be possible to act like a \textsc{mifare} reader to communicate with the card. The Proxmark III developed by Jonathan Westhues has these possibilities\footnote{Hardware design and software is publicly available at \url{http://cq.cx/proxmark3.pl}}. Because of its flexible design, it is possible to adjust the Digital Signal Processing to support a specific protocol. This device supports both low frequency (125\,kHz - 134\,kHz) and high frequency (13.56\,MHz) signal processing. The signal from the antenna is routed through a Field Programmable Gate Array (FPGA). This FPGA relays the signal to the microcontroller and can be used to perform some filtering operations before relaying.
\begin{figure}[ht]
\begin{center}
\includegraphics[width=90mm]{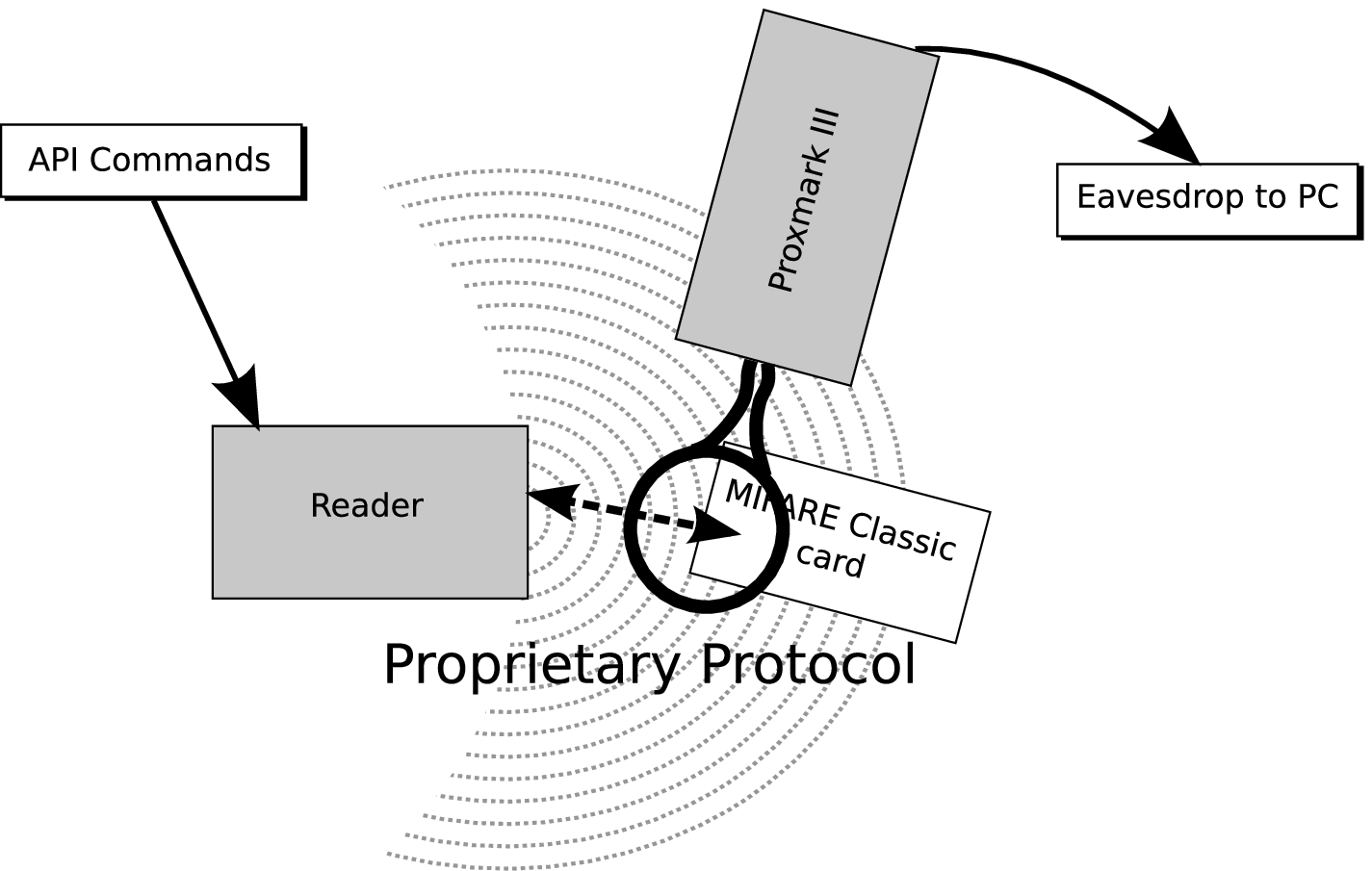}
\vspace{-15pt}
\caption{Experimental Setup}
\vspace{-25pt}
\label{picsetup}
\end{center}
\end{figure}
The software implementation allows the Proxmark to eavesdrop communication (Figure~\ref{picsetup}) between an RFID tag and a reader, emulate a tag and a reader. In this case our tag will be the \textsc{mifare} Classic card.
Despite the basic hardware support for these operations the actual processing of the digitized signal and (de)modulation needs to be programmed for each specific application. The physical layer of the \textsc{mifare} Classic card is implemented according to the ISO14443-A standard \cite{ISO14443}. We had to implement the ISO14443-A functionality since it was not implemented yet. This means we had to program both processing and generation of reader-to-tag and tag-to-reader communication in the physical layer and higher level protocol. To meet the requirements of a replay attack we added the functions `hi14asnoop' to make traces, `hi14areader' to act like a reader and `hi14asim' to simulate a card. We added the possibility to send custom parity bits. This was needed because parity bits are part of the encryption.

\section{Communication Characteristics}\label{comchar}
To find out what the \textsc{mifare} Classic communication looks like we made traces of transactions between \textsc{mifare} readers and cards. This way, we gathered many traces which gave us some insights on the high-level protocol of \textsc{mifare} Classic. In this section we explain a trace we recorded as an example, which is shown in Figure \ref{trace04}.
\begin{figure}[th]
$
\begin{array}{rrrll}
&\;\;\;\;\;\;\;\;\mathtt{ETU} & \mathtt{SEQ} & \;\; \mathtt{sender} & \mathtt{bytes}
\end{array}
$\\
\begin{tabular}{ll} \hline
$\begin{array}{lrlll}
& \mathtt{0:} & \mathtt{01 :} & \; \mathtt{PCD} \;\;\;\;\; & \mathtt{26}\\
& \mathtt{64:} & \mathtt{02 :} & \; \mathtt{TAG} & \mathtt{04\;\;00}\\
& \mathtt{12097:} & \mathtt{03 :} & \; \mathtt{PCD} & \mathtt{93\;\;20}\\
& \mathtt{64:} & \mathtt{04 :} & \; \mathtt{TAG} & \mathtt{2a\;\;69\;\;8d\;\;43\;\;8d}\\
& \;\;\;\;\mathtt{16305:} & \mathtt{05 :} & \; \mathtt{PCD} & \mathtt{93\;\;70\;\;2a\;\;69\;\;8d\;\;43\;\;8d\;\;52\;\;55}\\
& \mathtt{64:} & \mathtt{06 :} & \; \mathtt{TAG} & \mathtt{08\;\;b6\;\;dd}\\
\end{array}$ & $\left.\begin{array}{l}\\\\\\\\\\\\\end{array}\right\}$ Anticollision\\\hline
$\begin{array}{lrlll}
& \;\;\;\;\mathtt{16504:} & \mathtt{07 :} & \; \mathtt{PCD} \;\;\;\;\; & \mathtt{60\;\;04\;\;d1\;\;3d}\\
& \mathtt{112:} & \mathtt{08 :} & \; \mathtt{TAG} & \mathtt{3b\;\;ae\;\;03\;\;2d}\\
& \mathtt{6952:} & \mathtt{09 :} & \; \mathtt{PCD} & \mathtt{c4!\;94\;\;a1\;\;d2\;\;6e!\;96\;\;86!\;42}\\
& \mathtt{64:} & \mathtt{10 :} & \; \mathtt{TAG} & \mathtt{84\;\;66!\;05!\;9e!}\\
\end{array}$ & $\left.\begin{array}{l}\\\\\\\\\end{array}\right\}$ Authentication\\\hline
$\begin{array}{lrlll}
& \;\;\mathtt{396196:} & \mathtt{11 :} & \; \mathtt{PCD} \;\;\;\;\; & \mathtt{a0\;\;61!\;d3!\;e3}\\
& \mathtt{208:} & \mathtt{12 :} & \; \mathtt{TAG} & \mathtt{0d}\\
& \mathtt{8442:} & \mathtt{13 :} & \; \mathtt{PCD} & \mathtt{26\;\;42\;\;ea\;\;1d\;\;f1!\;68!}\\
& \mathtt{5120:} & \mathtt{14 :} & \; \mathtt{PCD} & \mathtt{8d!\;ca\;\;cd\;\;ea}\\
& \mathtt{2816:} & \mathtt{15 :} & \; \mathtt{TAG} & \mathtt{06!}\\
\end{array}$ & $\left.\begin{array}{l}\\\\\\\\\\\end{array}\right\}$ Increment \& Transfer\\\hline
$\begin{array}{lrlll}
& \mathtt{1349238:} & \mathtt{16 :} & \; \mathtt{PCD} \;\;\;\;\; & \mathtt{2a\;\;2b\;\;17\;\;97}\\
& \mathtt{72:} & \mathtt{17 :} & \; \mathtt{TAG} & \mathtt{49!\;09!\;3b!\;4e!\;9e!\;5e\;\;b0\;\;06\;\;d0!}\\
 & & & & \mathtt{07!\;1a!\;4a!\;b4!\;5c\;\;b0!\;4f\;\;c8!\;a4!}\\
\end{array}$ & $\left.\begin{array}{l}\\\\\\\end{array}\right\}$ Read\\\hline
\end{tabular}

\caption{Trace of a card with default keys, recorded by the Proxmark III}
\label{trace04}
\end{figure}
This trace contains every part of a transaction. We refer to the sequence number (SEQ) of the messages we discuss. The messages from the reader are shown as PCD (Proximity Coupling Device) messages and from the card as TAG messages. The time between messages is shown in Elementary Time Units (ETU). One ETU is a quarter of the bit period, which equals 1.18\,$\upmu$s. The messages are represented in hexadecimal notation. If the parity bit of a byte is incorrect\footnote{encrypted parity bits show up as parity error in the message}, this is shown by an exclamation mark. We will discuss only the most significant messages.
\paragraph{Anticollision} The reader starts the SELECT procedure. The reader sends \texttt{93 20} (\#3), on which the card will respond with its unique identifier (\#4). The reader sends \texttt{93 70} followed by the UID and two CRC bytes (\#5) to select the card.
\paragraph{Authentication}
The card is in the active state and ready to handle any higher layer commands.
In Section \ref{authentication} we discussed the authentication protocol. In Figure \ref{trace04}, messages \#7 to \#10 correspond to the authentication.\\
The authentication request of the reader is \texttt{60 04 d1 3d} (\#07). The first byte \texttt{60} stands for an authentication request with key A. For authentication with key B, the first byte must be \texttt{61}. The second byte indicates that the reader wants to authenticate for block 4. Note that block 4 is part of sector 1 and therefore this is an authentication request for sector 1. The last two bytes are CRC bytes.
\paragraph{Encrypted Communication}
After this successful authentication the card is ready to handle commands for sector 1. The structure of the commands can be recognized clearly. Since we control the \textsc{mifare} Classic reader we knew which commands were sent. Message \#11 to \#15 show how an \textit{increment} is performed. The \textit{increment} is immediately followed by a \textit{read} command (\#16 and \#17).

\section{Weakness in MIFARE Classic} \label{weakness}
Nohl and Pl\"otz partially recovered the CRYPTO1 algorithm that is used to encrypt the communication between the card and the reader~\cite{CCC,USENIX08}. The pseudo-random generator on the card, which initiates the algorithm by generating a nonce, is weak. In our analysis, we use this weakness to extend the work of Nohl and Pl\"otz with a practical attack, which delivers the needed known plaintext for brute-force, and a description of the \textsc{mifare} Classic protocol. In this attack, we do not need knowledge about the CRYPTO1 algorithm other than that it is a stream cipher which encrypts bitwise.
\paragraph{}
During our experiments, independently, we also noted the weakness of the pseudo-random generator of the card by requesting many card nonces. We were able to request about 600,000 nonces every hour. Within one hour, a nonce reappeared at least about four times. The nonce is generated by a \textit{Linear Feedback Shift Register} (LFSR)~\cite{USENIX08} which shifts every 9.44\,$\upmu$s. This is exactly one bit period in the communication. Therefore a random nonce could theoretically reappear after 0.618s, if the card is queried at exactly the right time.\\
In another experiment, we tried to request a nonce at a fixed time after powering-up\footnote{as was suggested by Nohl and Pl\"otz~\cite{CCC}} the card. This way, we could reduce the card nonces to ten different ones, which decreases the waiting time.
\paragraph{} 
Without knowing the cryptographic algorithm, only an online brute force attack on the key can be mounted. Because of the communication delay, this would take 5ms for each attempt. An exhaustive key search would then take 16,289,061 days, which equals about 44,627 years.\\
When the cryptographic algorithm is known, an off-line brute force attack can be mounted using a few eavesdropped traces of an authentication run. Nohl and Pl\"otz state that with dedicated hardware of around \$17,000 this would take about one hour. For this attack to work, some known plaintext is required. Our analysis provides this plaintext.

\section{Keystream Recovery Attack} \label{attack}
In Section \ref{weakness} we discussed a weakness in the pseudo-random generator of the \textsc{mifare} Classic. In this section we deploy a method to recover the keystream that was used in an earlier recorded transaction between a reader and a card. As a result the keystream of the communication will be recovered. For this attack we need to be in possession of the card. The following reasons make this attack interesting:
\begin{enumerate}
\item Our attack provides the known plaintext necessary to mount a brute force attack on the key.
\item Using our attack we recovered details about the byte commands.
\item Using the recovered keystream we can \textit{read} card contents without knowing the key.
\item Using the recovered keystream we can also \textit{modify} the contents of the card without knowing the key.
\end{enumerate}

\subsection{Keystream Recovery} \label{knowledge}
To recover the keystream we exploit the weakness of the pseudo-random generator. As it is this random nonce in combination with only one valid response of the reader what determines the remaining keystream. For this attack we need complete control over the reader (Proxmark) and access to a (genuine) card. The attack consists of the following steps:
\begin{enumerate}
\item Eavesdrop the communication between a reader and a card. This can be for example in an access control system or public transport system.
\item Make sure that the card will use the same keystream as in the recorded communication. This is possible because the card repeats the same nonce in reasonable time, and we completely control the reader.
\item Modify the plaintext, such that the card receives a command for which we know plaintext in the response (e.g., by changing the block number in a read command).
\item For each segment of known plaintext, compute the corresponding keystream segment.
\item Use this keystream to partially decrypt the trace obtained in 1.
\item Try recovering more keystream bits by shifting commands.
\end{enumerate}
\paragraph{}
The plaintext $P_{1}$ in the communication is XOR-ed bitwise with a keystream $K$ which gives the encrypted data $C_{1}$.
When it is possible to use the same keystream on a different plaintext $P_{2}$ and either $P_{1}$ or $P_{2}$ is known, then both $P_{1}$ and $P_{2}$ are revealed.
\begin{equation}
	\left.
	\begin{array}{l}
	P_{1} \oplus K = C_{1} \\
	P_{2} \oplus K = C_{2}
	\end{array}
	\right\}
	C_{1} \oplus C_{2} \Rightarrow P_{1} \oplus P_{2} \oplus K \oplus K \Rightarrow P_{1} \oplus P_{2}
\end{equation}
The weak pseudo-random generator makes it possible to replay an earlier recorded transaction. We can flip ciphertext bits to try to modify the first command such that it gives another result. Another result gives us another plain text. The attack is based on this principle.

\subsection{Keystream Mapping}
The data is encrypted bitwise. When the reader sends or receives a message, the keystream is shifted the number of bits in this message on both the reader and card side. This is needed to stay synchronized and use the same keystream bits to encrypt and decrypt. The stream cipher does not use any feedback mechanism. Despite that, when we tried to reveal the contents of a message sequence using a known keystream of an earlier trace, something went wrong.
We recorded an \textit{increment} followed by a \textit{transfer} command. We used this trace to apply our attack and changed the first command to a \textit{read} command which consists of 4 command bytes and delivers 18 response bytes. Together with the parity bits this makes it a 198 bit stream. The plaintext was known and therefore we recovered 198 keystream bits.

When we used this keystream to map it on the original trace of the \textit{increment} (Figure~\ref{trace03}), it turned out that the keystream was not in phase after the first command. The reason was the short 4-bit answer of the card that is not followed by a parity bit. In our original trace we are now half way the first response byte. This means that after 4 more bits we arrive at the parity bit in the original trace. However, in our new trace we are then half way the next command byte. To correct this we needed to throw away the keystream bit that was originally used to encrypt the parity bit.\\
But what to do when we need to decrypt a parity bit in the new situation and we are half way a byte with respect to the first trace? The solution is to encrypt the parity bit with the next bit from the recovered keystream and use this same keystream bit to decrypt the next data bit.\\
From this we can conclude that parity bits are encrypted with keystream bits that are also used to encrypt databits.
\begin{figure}[ht]
\begin{center}
\texttt{
\begin{tabular}{|l|l|l|l|l|l|l|}
\hline
         & INCREMENT & ACK & VALUE & TRANSFER & ACK\\\hline
Plaintext & c1 04 f6 8b & 0a & 01 00 00 00 bb 4a & b0 04 ea 62 & 0a\\\hline
Ciphertext   & 4c 88 31 bc! & 0a! & e2 79!2a!14 35!6f! & 04!81 2d!1e! & 0c!\\\hline
\end{tabular}
}
\caption{Recovering the Keystream and Commands}
\vspace{-20pt}
\label{trace03}
\end{center}
\end{figure}
\paragraph{}
The following method successfully maps the keystream on another message sequence as we described above.\\
Take the recovered keystream and strip all the keystream bits from it that were at parity bit positions. The remaining keystream can be used to encrypt new messages. Every time a parity bit needs to be encrypted, use the next keystream bit without shifting the keystream, in all other cases use the next keystream bit and shift the keystream.

\subsection{Authentication Replay} \label{replay}
To replay an authentication we first need a trace of a successful authentication between a genuine \textsc{mifare} reader and card. An example of an authentication followed by one read command is shown below.
\begin{center}
\begin{verbatim}
  1  PCD   60  03  6e  49
  2  TAG   e0  92  93  98
  3  PCD   ad  e7  96! 48! 20! 22  df  93
  4  TAG   bf  06  91! 82
  5  PCD   b5! 05! 47  3f
  6  TAG   3f  14! 4f  e9! 86  38! 96! 85 3e!
           f3  e3! 3d! eb! 2b! a2  d4  dd 76!
\end{verbatim}

\label{trace01}
\end{center}
After we recorded an authentication between card and reader, we do not modify the memory. This ensures that the memory of the card remains unaltered and therefore it will return the same plaintext. Now we will act like a \textsc{mifare} reader and try to initiate the same authentication. In short:
\begin{enumerate}
\item We recorded a trace of a successful authentication between a genuine card and reader.
\item We send authentication requests (\#1) until we get a nonce that is equal to the one (\#2) in the original trace.
\item We send the recorded response (\#3) to this nonce. It consists of a valid response to the challenge nonce and a challenge from the reader.
\item We retrieve the response (\#4) to the challenge from the card.
\item Now we are at the point where we could resend the same command (\#5) or attempt to modify it.
\end{enumerate}

\subsection{Reading Sector Zero}
We will show that it is possible to read sector 0 from a card without knowing the key. We only need one transaction between a genuine \textsc{mifare} reader and card. Every \textsc{mifare} Classic card has some known memory contents. The product information published by NXP \cite{NXPMIFARE4K} gives this information.\\
When a sector trailer is read the card will return logical `0's instead of key A because key A is not readable. If key B is not readable the card also returns logical `0's. It depends on the access conditions if key B is readable or not.
\vspace{-10pt}
\begin{figure}[ht]
\begin{center}
\includegraphics[width=70mm]{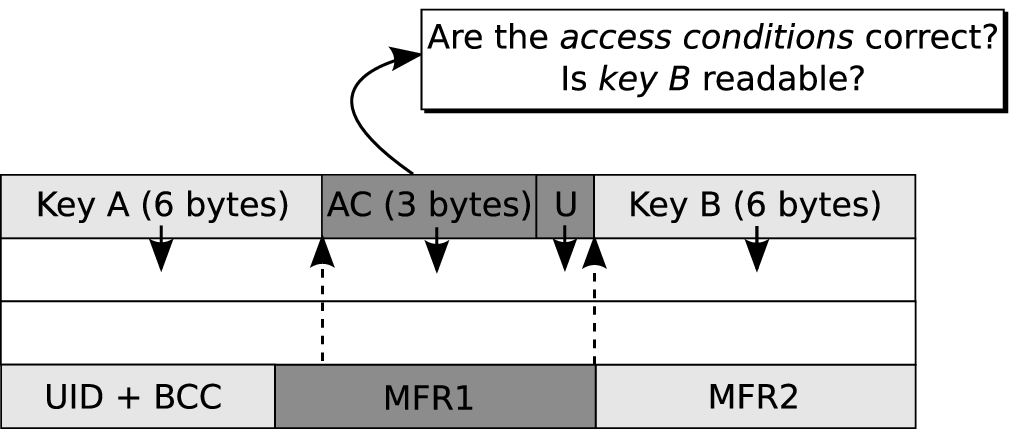}
\caption{Recovering Sector Zero}
\label{picsector0}
\vspace{-30pt}
\end{center}
\end{figure}
The access conditions can be recovered by using the manufacturer data. Block 0 contains the UID and BCC followed by the manufacturer data. The UID and BCC cover 5 bytes and are known. The remaining 11 bytes are covered by the manufacturer data. Some investigation on different cards (\textsc{mifare} Classic 1k and 4k) revealed that the first 5 bytes of the manufacturer data almost never change. These bytes (MFR1) cover the positions of the access conditions (AC) and the unknown byte U, as shown in Figure \ref{picsector0}. This means that the keystream can be recovered using the known MFR1 bytes by reading block 0 and block 3 (sector trailer) subsequently. Remember that the access conditions are stored twice in 3 bytes. Once inverted and once non-inverted. This way it is easy to detect if we indeed revealed the access conditions. The unknown byte U can be in any state when the card leaves the manufacturer but appears to be often \texttt{00} or \texttt{69}.

The access conditions tell us whether key B is readable or not. In many cases key B is not readable, for instance as in the OV-Chipkaart\footnote{\textsc{mifare} Classic 4k card} that is used in the Dutch public transport system. The first 5 bytes of the manufacturer data (MFR1 in Figure \ref{picsector0}) recovered the access conditions for sector 0. Because the access conditions for the sector trailer define key B as not readable, we know the plaintext is zeros. Hence the whole sector trailer was revealed and therefore the contents of the whole sector 0 were revealed as well.

\section{Reading Higher Sectors}
In the higher sectors of the \textsc{mifare} Classic card we do not have the advance of the manufacturer data. We basically have the sector trailer and some unknown data blocks. Because of key A we can recover always the first 10 keystream bytes. Key B is in most cases not readable and therefore will give 6 more keystream bytes, but leaves us with a gap of 4 bytes (AC and U).\\
Although it is harder to achieve, there is a potential threat for these sectors to become compromised.

\subsection{Proprietary Command Codes}
At the time this research was performed, we were not aware that the command codes, which we revealed with our attack, could already be found in example firmware of NXP\footnote{\url{http://www.nxp.com/files/markets/identification/download/MC081380.zip}}. Note that the firmware refers to the command codes sent from PC to reader. Our research shows that (perhaps obviously) these are the same command codes sent from reader to card. 

\begin{figure}[hb]
\begin{center}
\includegraphics[width=80mm]{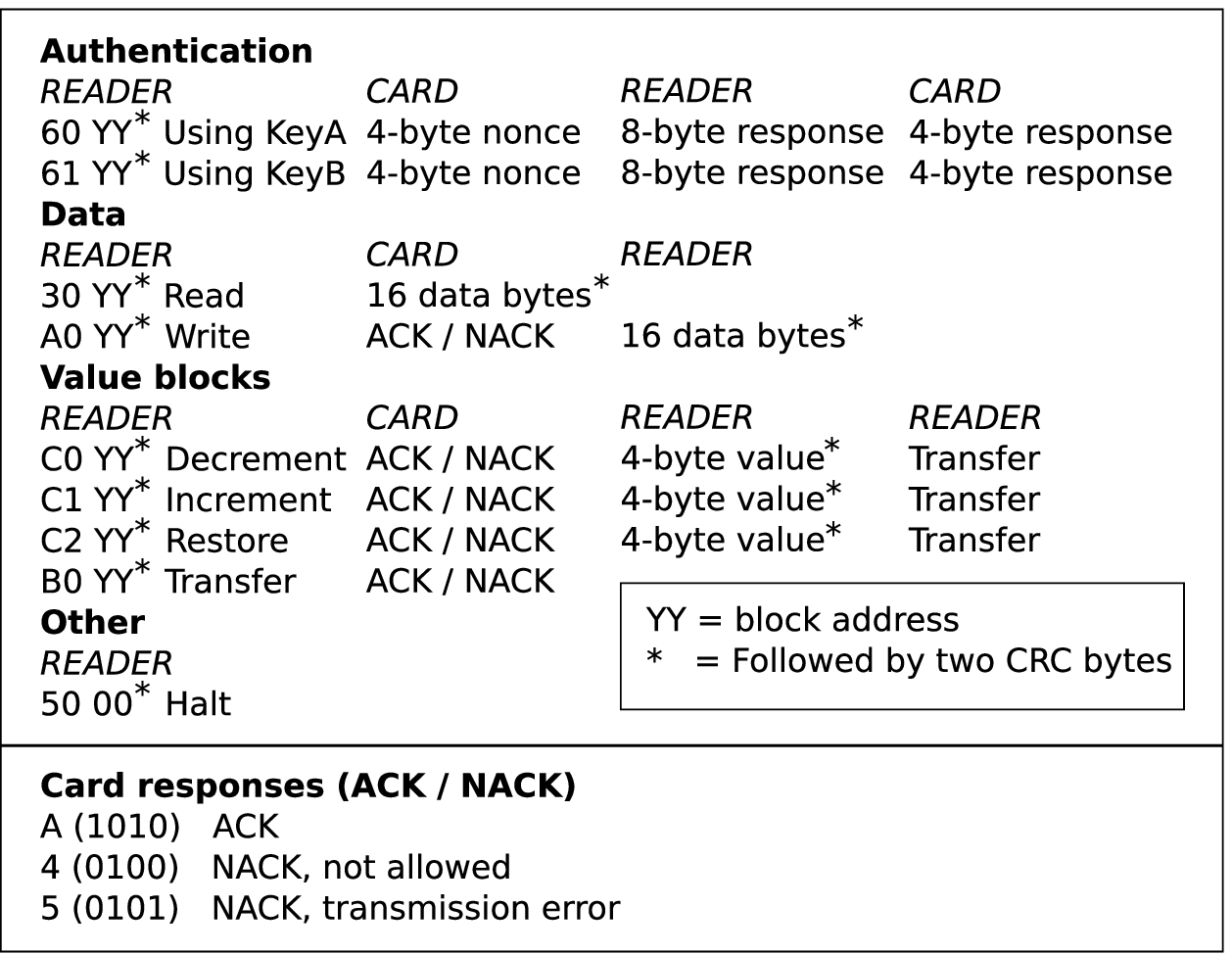}
\caption{Command set of \textsc{mifare} Classic}
\label{tabcommands}
\vspace{-15pt}
\end{center}
\end{figure}

We used a card in transport configuration with default keys and empty data blocks to reveal the encrypted commands used in the high-level protocol.
All the commands send by the reader consist of a command byte, parameter byte and two CRC bytes. We made several attempts to reveal the command by modifying the ciphertext of this command. The way to do this is to assume we actually know the command. With this `knowledge' we XOR the ciphertext which gives us the keystream. To check if this is indeed the correct keystream, we XOR it with a new command for which we know the response. If we guessed the initial command right the response of the card will be that known response. This method revealed the commands shown in Figure~\ref{tabcommands}.

Now, one could try to replay the same authentication again and try to execute a command that returns an ACK or NACK in order to recover more keystream. Because an ACK or NACK is only 4 bits in size, it leaves some spare bits for which we know the keystream. We can use these bits to execute another command for which we now know the plaintext. This delivers more known keystream as a result, and this method can be applied repeatedly. However, this approach does only work if a \textit{decrement}, \textit{increment} or \textit{transfer} is allowed. These are the commands that return an ACK and therefore are in total shorter than the \textit{read}. We can only send valid commands because otherwise the protocol aborts.
The \textit{read} command returns 16 data bytes and 2 CRC bytes. On a \textit{write} command the card returns a 4-bit ACK, this indicates that the card is ready to receive 16 data bytes followed by 2 CRC bytes.\\
The \textit{decrement}, \textit{increment} and \textit{restore} commands all follow the same procedure. The card indicates that it is expecting a value from the reader by sending a 4-bit ACK response. This value is 4 bytes and is followed by 2 CRC bytes. For the \textit{restore} this value is send but not used. The value is send as \texttt{YY YY YY YY ZZ ZZ}, where \texttt{YY} are the value bytes and \texttt{ZZ} the CRC bytes.\\
Finally, a \textit{transfer} command is send to transfer the result of one of the previous commands to a memory block. The card response is an ACK if it went well. Otherwise it responds with a NACK.

The 4-bit ACK is \texttt{0xa}. When a command is not allowed the card sends \texttt{0x4}. When a transmission error is detected the card sends \texttt{0x5}. The card does not even give a response at all if the command is of the wrong length. The protocol aborts on every mistake or disallowed command.

\section{Conclusions \& Recommendations} \label{conclusions}

We have implemented a successful attack to recover the keystream of an earlier
recorded transaction between a genuine \textsc{mifare} Classic reader and card.

We used a \textsc{mifare} Classic reader in combination with a `blank' card
with default keys to recover the byte commands that are used in the proprietary
protocol.
Knowing the byte commands and a sufficiently long keystream allowed us to
perform any operation as if we were in possession of the secret key.

We managed to read \emph{all} memory blocks of the sector zero of the card, without having access to the secret key. In general, we were able to read \emph{any} sector of the memory of the card, provided that we know \emph{one} memory block within this sector. Moreover, after recording a valid transaction on any sector, we were able to read the first 6 bytes of any block in that sector and also the last 6 bytes if key B is read only.
Similarly, we are able to \emph{modify} the information stored in a particular
sector.

\paragraph{Consequences}

First of all, all data stored on the card (except the keys themselves) should
no longer be considered secret. In particular, if the \textsc{mifare} Classic card is used to
store personal information (like name, date of birth, or travel information),
this constitutes a direct privacy risk. The security risk is relatively low
because in general the security is guaranteed by the secrecy of the keys.
Note that in particular we are not able to clone cards,
because the secret keys remain secret.

Secondly, the integrity and authenticity of the data stored on the card can no
longer be relied on. This is quite a severe security risk. This is particularly
worrying in applications where the card is used to store a certain value, like
loyalty points or, even worse, some form of digital currency. The loyalty level
or the value stored in the electronic purse could easily be increased (or
decreased, in a denial-of-service type of attack).

Thirdly, knowledge of the plaintext (or the keystream) is a necessary
condition to perform brute force (or other more sophisticated) attacks to
recover the secret key. We are making good progress in developing a very
efficient attack to recover arbitrary sector keys of a \textsc{mifare} Classic card.


\paragraph{Recommendations}

For short term improvements we recommend not to use sector zero to store secret information. Configure key B as readable and store random information in it. Do not store sensitive information in the first 6 bytes of any sector.
Use multiple sector authentications in one transaction to thwart attackers in an attempt to recover plaintext. This is only helpful when value block commands are not allowed. Value block commands are shorter than a read command and will enable a shift of the keystream. Another possibility, that might be viable for some applications, is to employ another encryption scheme like AES in the backoffice, and store only encrypted information on the tags.
To prevent unauthorized modification of a data block, an extra authentication on this data could be added. This authentication is then verified in the backoffice.

Proper fraud detection mechanisms and extra security features in the backoffice
are necessary to signal or even prevent the types of attacks described
above. In general, the backoffice systems collecting and processing data that
comes from the readers are a very important second line of defense.

On the long term these countermeasures will not be sufficient. The \textsc{mifare} Classic card has a closed design. Security by obscurity has shown several times that at some point the details of the system will be revealed compromising security~\cite{kerckhoffs1883cryptographie}. Therefore we recommend to migrate to more advanced cards with an open design architecture. 

\bibliographystyle{plain}
\bibliography{mifare_weakness}

\end{document}